\documentclass{amsart}
\def\curl{{\rm curl\,}}
\def\div{{\rm div\,}}
\def\R{\mathbb{R}}
\def\C{\mathbb{C}}

\def\HH{\mathcal{H}}
\def\KK{\mathcal{K}}
\def\e{\mathrm{e}}

\theoremstyle{plain}
\newtheorem{theorem}{Theorem}
\newtheorem{lemma}{Lemma}
\newtheorem{corollary}{Corollary}
\theoremstyle{remark}
\newtheorem*{remark}{Remark}

\begin{document}

\markboth{A.\ M.\ Alekseenko}
{Well-posed IBVP for a constrained evolution system and constraint-preserving BCs}

\title[Well-posed IBVP for a constrained evolution system and the BCs]
{Well-Posed Initial-Boundary Value Problem for a Constrained Evolution
System and Radiation-Controlling Constraint-Preserving Boundary
Conditions}

\author{Alexander M. Alekseenko}

\address{Department of Mathematics, California State University
Northridge, 18111 Nordhoff St., Northridge, CA 91330-8313}

\email{alexander.alekseenko@csun.edu}

\maketitle

\begin{center}
Submitted 5 Aug. 2006\\
Revised 21 Mar. 2007\\
\end{center}

\begin{abstract}
{\bfseries Abstract.}\quad A well-posed initial-boundary value
problem is formulated for the model problem of the vector wave
equation subject to the divergence-free constraint. Existence,
uniqueness and stability of the solution is proved by reduction to
a system evolving the constraint quantity statically, i.e., the
second time derivative of the constraint quantity is zero. A new
set of radiation-controlling constraint-preserving boundary
conditions is constructed for the free evolution problem.
Comparison between the new conditions and the standard
constraint-preserving boundary conditions is made using the
Fourier-Laplace analysis and the power series decomposition in
time. The new boundary conditions satisfy the Kreiss condition and
are free from the ill-posed modes growing polynomially in time.
\end{abstract}

\section{Introduction}

When formulated for the purpose of numerical integration the equations
of general relativity are customary divided into the two subsets:
the subset that contains both time and space derivatives, called
the evolution subset, and the subset that has spatial derivatives
only, called the constraint subset. Usually, the evolution subset
is the one that is being solved, while the constraint subset is
made automatically compatible by a choice of initial and boundary
data. Exact compatibility, however, can never be achieved in
numerical calculations, which are the primary source of solutions
to Einstein's equations. To predict accuracy of the solution, a
new formalism is needed to describe systems of constrained
evolution.

In fact, it has been a challenge in numerical relativity to keep
the constraint subset of equations from exponential growth
\cite{BFHR99,GMG04b,LSKPST04,SKLPT02}, when evolution equations
are solved exclusively. Methods were introduced to enforce
compatibility of the solution with the constraint equations by
adding differential terms to the evolution equations so as to
minimize growth of the constraints (e.g., \cite{LSKPST04}),
periodically resolving constraint equations, when they become too
large, (e.g., \cite{HLOPSK04}), and controlling the constraint
perturbations from entering the domain through the timelike
boundaries (e.g.,
\cite{A04,AT06,CLT02,CPST03,GMG04b,KLS04,LSKPST04,SW03,ST05}). The
method proposed in this paper is related to the last group. We
explore the idea of \cite{FN99} and study the closely related
system that propagates the constraint quantities statically. The
results are used to formulate a well-posed initial-boundary value
problem for the original constrained evolution system. On the next
step, the constrained problem is used to define a well-posed
initial-boundary value problem for the free evolution.

We consider a model problem of the vector wave equation
\begin{equation}
\label{eq1}
\partial^2_{t} u_{i}=\partial^{j}\partial_{j}u_{i},
\end{equation}
subject to the differential constraint
\begin{equation}
\label{eq2}
\partial^{i}u_{i}=0.
\end{equation}
The choice of the model problem is motivated by re-formulations of
Einstein's field equations (e.g.,
\cite{ADM,York,BS98,SN95,NOR04,GMG04a,GMG04b,RB04}) in which
second order in space equations are coupled to first order
differential constraints (if necessary, (\ref{eq1}) can be reduced
to first order in time by introducing $\pi_{i}=\partial_{t}u_{i}$;
all results can be repeated for the pair $(u_{i},\pi_{i})$ without
much complication). First order symmetric hyperbolic reductions of
(\ref{eq1}) has been used as a source of insight in numerical
relativity in the past \cite{C04,F04,LSKPST04,RS05} and methods
for constructing constraint-preserving boundary conditions were
investigated using the same model problem. In this work we prove
that (\ref{eq1}), (\ref{eq2}), taken as a \textit{constrained
evolution problem}, is related very closely to the equation
\begin{equation*}
\partial^2_{t} u_{i}=\partial^{j}\partial_{j}u_{i}-\partial_{i} \partial^{j}u_{j}
\end{equation*}
which evolves the constraint (\ref{eq2}) statically. Based on this
observation, we formulate an initial-boundary value problem for
the constrained evolution system (\ref{eq1}), (\ref{eq2}), and
prove existence, uniqueness, and stability of its solution.

The first application of the new approach is the new set of
radiative constraint-preserving boundary conditions (see e.g.,
\cite{LSKPST04,KLS04,RS05,ST05,KW06,R06}) for the free evolution
problem (\ref{eq1}). In the simplest case, when homogeneous
conditions are specified on the incoming waves, the boundary
conditions take the form
\begin{equation}
\label{eq3}
\partial_{t}u_{A} + \partial_{n}u_{A}-\partial_{A}u_{n} = 0,\quad
\partial_{t}u_{n} + \partial_{n}u_{n}=0,
\end{equation}
where $\partial_{n}$ stands for the normal, and $\partial_{A}$ for
the tangential derivatives at the boundary, and $u_{n}$, $u_{A}$
are projections of $u_{i}$ on the outward pointing unit normal and
the tangential directions, respectively.

In \cite{RS05}, an example of constraint-preserving
radiation-controlling boundary conditions is constructed for a
symmetric hyperbolic reduction of the fat Maxwell system. The
boundary conditions (\ref{eq3}) overlap with the boundary
conditions of \cite{RS05} in two, but differ significantly in one
equation. Following the standard approach of the subsidiary
evolution system for constraint quantities, the radiative boundary
condition enforcing the constraint in \cite{RS05} takes the form
of a first order differential condition (containing both spatial
and temporal derivatives) for the first order system. Here,
however, the new approach allows us to write all three conditions
as first order conditions (but still including tangential
derivatives) for the second order system, and existence is proved
without any additional gauge assumptions.

The problem (\ref{eq1}), (\ref{eq2}) has been considered in
\cite{KW06} in the case of a four-dimensional vector $u$, e.g.,
$u=(u_{t},u_{x},u_{y},u_{z})$. As in \cite{RS05}, the intuition
derived from Maxwell's equations has been used to formulate a set
of Sommerfeld-type boundary conditions that are
constraint-preserving. In fact, three of the four conditions of
\cite{KW06} coincide with (\ref{eq3}) if the component $u_{t}$ is
set to zero. The fourth condition, however, is closely related to
the homogeneous Dirichlet data for the constraint quantity. It
does not vanish for $u_{t}=0$, and, therefore, the boundary
conditions of \cite{KW06} do not reduce to (\ref{eq3}).

Several authors (e.g., \cite{KLS04, RB04}) had experimented with
boundary conditions motivated by physics rather than the equation
analysis. Surprisingly, the stability of the calculations improved
when physical data was employed. Their results motivated us to
search for the mathematical framework that could explain such
effect, and possibly describe the accurate boundary conditions for
constrained evolution systems.

The idea of reducing Einstein's equations to a form evolving the
constraint quantities statically was first used in \cite{FN99}. By
exploiting gauge freedom, the authors re-write the tetrad
formulation of Einstein's equations as to have the constraint
quantities propagating along the boundary (but not inside the
domain), thus making any boundary conditions that are well-posed
for the problem automatically compatible with preservation of the
constraints. This approach allowed the authors to formulate a set
of maximally dissipative boundary conditions for which
well-posedness can be proved in the full nonlinear case using
standard theorems for symmetric hyperbolic systems.

In this work we use constraint equations --- rather than gauge
freedom --- to formulate the static constraint evolution
reduction, as described in Sections~2 and 3. In Section~4 we use
standard theorems for first order symmetric hyperbolic systems to
prove the existence of the solution to the static constraint
evolution problem. In Section~5 we formulate a well-posed initial
boundary value problem for the constrained evolution system
(\ref{eq1}), (\ref{eq2}). The existence, uniqueness, and stability
of the solution follows from the theorems for the static
constraint evolution problem. We also formulate a set of
constraint-preserving radiative-controlling boundary conditions
for the free evolution problem (\ref{eq1}) and prove that this set
is well-posed. Notice that the norm in which the well-posedness is
proved is built on the antisymmetric part of the gradient and the
divergence of the solution. However we believe that this is the
natural norm for the problem, and any ``extra'' estimates on the
gradient require additional assumptions, like more restrictive
boundary conditions exemplified in Section~6, or a special gauge
choice similar to one in \cite{RS05}.

Section~7 is devoted to the Laplace-Fourier analysis of the free
evolution problem. We prove that the new boundary conditions
(\ref{eq3}) satisfy the Kreiss condition, thus the free evolution
problem is stable for zero initial data. (Notice that the stability of
the system can be studied for arbitrary initial data using
techniques of \cite{KL89,KW06} if the Kreiss condition is
satisfied). We find that the Neumann-Dirichlet
constrain-preserving conditions obtained by the standard method do
not meet the Kreiss condition. Also, the radiative differential
conditions following from the standard approach were found not to
meet the Kreiss condition. Finally in Section~8 we study stability
of the system in the case of non-vanishing initial data using the
framework developed in \cite{R06} to quantify instabilities that
grow polynomially in time. It was found that both the new
conditions (\ref{eq3}) and the radiative differential condition
similar to one in \cite{RS05} do not expose such instabilities,
while the standard Neumann-Dirichlet data does.

\section*{Notations}

We denote by $\Omega$ a bounded convex subset of $\R^3$ with
boundary consisting of parts of planes. The polyhedron domains are
relatively easy to implement numerically, and therefore represent
an important example from the practical point of view. Vector $n_{i}$
in most cases denotes the outward pointing unit normal vector to
$\partial \Omega$. Vectors $m_{i}$ and $l_{i}$ are tangential to
$\partial \Omega$ and complement $n_{i}$ to an orthonormal triple.

We use notations of general relativity, in which the repeated
indices denote summation, small roman indices run from $1$ to $3$,
the indices are raised and lowered with the flat metric
$\delta_{ij}=\{ 1\ \mbox{if}\ i=j,\ 0\ \mbox{if}\ i\neq j\}$.
Later we use parenthesised and bracketed indices to denote the
symmetric and anti-symmetric parts of an array correspondingly,
e.g., $w_{(ij)}=(w_{ij}+w_{ji})/2$,
$w_{[ij]}=(w_{ij}-w_{ji})/2$, $w_{ij}=w_{(ij)}+w_{[ij]}$.

The symbol $\partial_{i}$ denotes the gradient, however,
$\partial_{n}$ is reserved for the directional derivative in
$n_{i}$. We will use capital roman indices to indicate the
tangential components of the fields, for example, $\partial_{A}$
stands for derivatives in tangential directions to $\partial
\Omega$ (or, if $n_{i}$ is not associated with $\partial\Omega$,
$\partial_{A}$ will denote derivatives in directions normal to
$n_{i}$). Similar notations are used for vectors:
$v_{n}=n^{j}v_{j}$, $v_{A}  = (\delta^{j}_{i}-n_{i}n^{j}) v_{j}$.

\section{Subordinate constraint evolution and constraint preservation}

Our goal is to formulate a well-posed initial-boundary value
problem for the constrained evolution system (\ref{eq1}),
(\ref{eq2}). We introduce the constraint quantity
\begin{equation*}
C=\partial^{i}u_{i}.
\end{equation*}

\begin{lemma}
\label{lem1}
If $u_{i}$ is a solution of (\ref{eq1}) then the constraint quantity $C$
obeys the wave equation,
\begin{equation}
\label{eq9}
\partial^2_{t} C=\partial^{i}\partial_{i}C.
\end{equation}
\end{lemma}
\begin{proof} The lemma follows by taking the divergence of Eq.~(\ref{eq1}) and
commuting derivatives.
\end{proof}

\begin{corollary}
\label{cor1} Consider the problem of solving Eq.~(\ref{eq1}) in
$\R^3$. Let $u_{i}$ be a solution to (\ref{eq1}), corresponding to
constraint-compatible initial data, i.e.,
$C(0)=\partial^{i}u_{i}(0)=0$ and
$\partial_{t}C(0)=\partial^{i}\partial_{t}u_{i}(0)=0$, and $u_{i}$
is regular to guarantee existence of the spatial integrals of $C$
and its derivatives. Then, $C\equiv 0$.
\end{corollary}
\begin{proof} According to Lemma~\ref{lem1}, the constraint quantity $C$
is a solution to (\ref{eq9}). Multiply (\ref{eq9}) by
$\partial_{t}C$ and integrate over $\R^3$. After
integration by parts,
\begin{equation*}
\int_{\R^3}(\partial^2_{t}C )(\partial_{t}C )dx=
-\int_{\R^3}(\partial_{i}C)(\partial^{i}\partial_{t}C) dx.
\end{equation*}
By commuting derivatives and re-grouping terms, we obtain
\begin{equation*}
\frac{1}{2}\partial_{t}[
\|\partial_{t}C\|^2_{L_{2}(\R^3)}+\|\partial_{i}C\|^2_{L_{2}(\R^3)}]
=0.
\end{equation*}
Therefore,
$\|\partial_{t}C\|^2_{L_{2}(\R^3)}+\|\partial_{i}C\|^2_{L_{2}(\R^3)}$
is constant in time. Since it is zero initially, it remains
zero for all times, which in turn gives that $C\equiv 0$.
\end{proof}

\begin{remark} Note that Corollary~\ref{cor1} does not hold if
$\R^3$ is replaced with a bounded region. Indeed, if $C\neq 0$ at
some point of the boundary, in view of (\ref{eq9}), the constraint
perturbation will flow inside the domain regardless of the
compatibility of the initial data.

In general to ensure $C\equiv 0$ the boundary data must be given
consistently with the evolution of the constraint quantity. The
data that guarantees that $C$ remains zero for all times is called
\textit{constraint-preserving}.
\end{remark}

\section{Static constraint evolution reduction}

In domain $\Omega$, consider the equation
\begin{equation}
\label{eq11}
\partial^2_{t} u_{i}=\partial^{j}\partial_{j}u_{i}-\partial_{i} C
\end{equation}
subject to constraint (\ref{eq2}).

\begin{lemma}
\label{lem2} Evolution of the constraint quantity $C$ in
Eq.~(\ref{eq11}) is static. That is, let $u_{i}$ be any
(sufficiently smooth) solution to (\ref{eq11}), then
\begin{equation}
\label{eq12}
\partial^{2}_{t}C=0.
\end{equation}
\end{lemma}

\begin{proof}
The proof follows by substituting the definition of $C$, commuting derivatives, and
using (\ref{eq11}).
\end{proof}

\begin{remark}
One may notice that in (\ref{eq11}), we recover the second
order reduction of Maxwell's equations in the Coulomb gauge. Indeed, by substituting the
definition of $C$, we rewrite (\ref{eq11}) as
\begin{equation*}
\partial^2_{t} u_{i}=2\partial^{j}\partial_{[j}u_{i]}=(\curl\,\curl
u)_{i},
\end{equation*}
where $(\curl u)_{i}$ is the usual curl operator defined by
$(\curl u)_{i}=\varepsilon^{\phantom{i}jk}_{i}\partial_{j}u_{k}$;
here $\varepsilon_{ijk}$ is the unit totally antisymmetric tensor,
and we have made use of the identity
$\varepsilon_{pjk}\varepsilon^{pmn}=
2\delta^{m}_{[j}\delta^{n}_{k]}$.

It is known that the divergence constraint
is preserved trivially for Maxwell's equations.
\end{remark}

Using notations of \cite{T96}, we introduce the linear algebraic operator
$L^{\hspace{2mm}pq}_{ji}=2\delta^{p}_{[j}\delta^{q}_{i]}$ acting on matrix
fields in $\Omega\subset\R^3$, and rewrite Eq.~(\ref{eq11})
as
\begin{equation}
\label{eq13}
\partial^2_{t} u_{i}=\partial^{j}L_{ji}^{\hspace{2mm}pq}\partial_{p}u_{q}.
\end{equation}

\begin{lemma}
\label{lem3}
The operator $L^{\hspace{2mm}pq}_{ji}$ is symmetric in the
sense of the scalar product
\begin{equation*}
\langle \phi_{ij},\psi_{ij} \rangle := \int_{\Omega}
\phi_{ij}\psi^{ij} dx = \int_{\Omega}
\phi_{ij}\psi_{ab}\delta^{a i}\delta^{b j} dx.
\end{equation*}
Let vectors $n_{i}$, $m_{i}$, and $l_{i}$ constitute an
orthonormal triple in $\R^3$. The eigenvalues and eigenvectors of
$L^{\hspace{2mm}pq}_{ji}$ are given by
\begin{align*}
2:\quad & \sqrt{2}\,n_{[j}m_{i]},\ \sqrt{2}\,n_{[j}l_{i]},\ \sqrt{2}\,m_{[j}l_{i]} \\
0:\quad & \sqrt{2}\,n_{(j}m_{i)},\ \sqrt{2}\,n_{(j}l_{i)},\ \sqrt{2}\,m_{(j}l_{i)},\
n_{j}n_{i},\ m_{j}m_{i},\  l_{j}l_{i}.
\end{align*}
\end{lemma}

\begin{proof}
The symmetry property follows by direct substitution and rearrangement of
dummy indices:
\begin{gather*}
\langle \phi_{ji}, L^{\hspace{2mm}pq}_{ji}\psi_{pq} \rangle =
\int_{\Omega} \phi_{ji}L^{\hspace{2mm}pq}_{ab}\psi_{pq}\delta^{aj}\delta^{bi}=
\int_{\Omega} \phi_{ji}[\delta^{pj}\delta^{qi}-\delta^{pi}\delta^{qj}]\psi_{pq}\\
{}=\int_{\Omega} \phi_{ji}[\delta^{j}_{a}\delta^{i}_{b}-\delta^{j}_{b}\delta^{i}_{a}]\psi_{pq}
\delta^{a p}\delta^{b q}
=\int_{\Omega} \phi_{ji}L^{\hspace{2mm}ji}_{ab} \psi_{pq}\delta^{a p}\delta^{b q}=
\langle L^{\hspace{2mm}ji}_{pq}\phi_{ji}, \psi_{pq} \rangle.
\end{gather*}
Note also that in the standard basis for $3\times 3$ matrices, $\mathbf{e1}=n_{i}n_{j}$, $\mathbf{e2}=n_{i}m_{j}$,
$\mathbf{e3}=n_{i}l_{j}$, \ldots, $\mathbf{e9}=l_{i}l_{j}$, the operator $L^{\hspace{2mm}pq}_{ji}$
takes the form of a $9\times 9$ symmetric matrix.
\end{proof}

\begin{remark}
Spectral structure of $L^{\hspace{2mm}pq}_{ji}$ suggests that
Eq.~(\ref{eq13}) does not allow control of the components of
$\partial_{p}u_{q}$ that are spanned by the zero eigenvectors,
since $L^{\hspace{2mm}pq}_{ji}$ annihilates them. However, this
conclusion is often only partly true. An energy estimate on
the entire gradient can be established using methods described in
Section~6.
\end{remark}

\section{First order reduction and boundary conditions}
Well-posedness of (\ref{eq13}), (\ref{eq2}) is proved by the first
order symmetric hyperbolic reduction. We introduce the components
of $\partial_{j}u_{i}$ spanned by the nonzero eigenvectors
of $L^{\hspace{2mm}pq}_{ji}$ as the new variables, namely,
$\psi_{ji}=2\partial_{[j}u_{i]}$. Then Eq.~(\ref{eq13}) is reduced to
\begin{align}
\partial_{t}u_{i}&=\pi_{i},\nonumber \\
\partial_{t}\pi_{i}&=\partial^{j}\psi_{ji},\nonumber \\
\label{eq14}
\partial_{t}\psi_{ji}&=2\partial_{[j}\pi_{i]}.
\end{align}
The initial data for the new variables is given by
\begin{equation}
\label{eq14a}
\pi_{i}(0)=\partial_{t} u_{i}(0),\qquad
\psi_{ji}(0)=2\partial_{[j}u_{i]}(0).
\end{equation}

\begin{remark}
Equation $\psi_{ji}=2\partial_{[j}u_{i]}$, the definition of
variable $\psi_{ji}$, formally has to be added to (\ref{eq14}) as
an extra (artificial) constraint equation to guarantee the
equivalence of (\ref{eq14}) and (\ref{eq13}). However, this
constraint follows from the last equation of (\ref{eq14}) by
integration in time. Because of the simple behaviour, we postpone it in
the analysis.
\end{remark}
\begin{remark}
A different first order decomposition, corresponding to
$\tilde{\psi}_{ji}=\partial_{j}u_{i}$,
$\pi_{i}=\partial_{t}u_{i}$, can be proposed for (\ref{eq13}). The
resulting system is called the fat Maxwell system and is only
weakly hyperbolic. A detailed study of this system can be found in
\cite{RS05}. However, Lemma~2 predicts that the symmetric part of
the variable $\tilde{\psi}_{ji}$ is not ``seen'' by the right side
of (\ref{eq13}). This results in the incomplete set of
characteristic variables for the fat Maxwell system. It is
possible, however, to obtain a strongly hyperbolic system by
adding a suitable combination of constraints to the evolution
equations.
\end{remark}

Let us introduce the scalar product for vectors fields, $\langle
u_{i},v_{i} \rangle=\int_{\Omega} u^{i}v_{i} dx$, and the norms
$\| u \|^2_{L_2(\Omega)}=\langle u_{i},u_{i} \rangle$, $\| \psi \|^2_{L_2(\Omega)}=\langle
\psi_{ij},\psi_{ij} \rangle$. By contracting the last two
equations of (\ref{eq14}) with $\pi^{i}$ and $\psi^{ji}$
correspondingly, integrating by parts, and re-grouping terms, we
have
\begin{equation}
\label{eq15}
\partial_{t}[2\|\pi\|^2_{L_2(\Omega)}+\|\psi\|^2_{L_2(\Omega)}] = 4\int_{\partial \Omega}
n_{[j}\pi_{i]}\psi^{ji}d \sigma.
\end{equation}

The boundary conditions for system (\ref{eq14}) are obtained from
the theory of symmetric hyperbolic systems. Let $n_{i}$
be the outward unit normal to $\partial \Omega$. Then the
characteristic speeds and fields in the direction $n_{i}$ are
\begin{align*}
    0:\quad     & \pi_{n},\quad \psi_{AB},\\
\pm 1:\quad     & w_{A}^{\pm}=\pi_{A} \pm \psi_{nA}.
\end{align*}
On each face of the boundary $\partial \Omega$, we impose the
maximal non-negative boundary conditions (in the sense of \cite{LP60,LP73,R85,M75}, see also in \cite{LSKPST04,RS05}),
\begin{equation}
\label{eq16}
w_{A}^{+}= \alpha^{B}_{A} w_{B}^{-}+g_{A},
\end{equation}
where the smooth matrix $\alpha^{B}_{A}$ is defined on $\partial
\Omega$, and satisfies $\|\alpha\|=\sup
\frac{\|\alpha^{B}_{A}v_{B}\|}{\| v \|}\le 1$. Since one may think
of the characteristic variables as the coefficients of the zero,
incoming, and outgoing modes correspondingly, condition
(\ref{eq16}) relates the amount of incoming radiation to the
outgoing.

Let us consider the case of homogeneous data first. If $g_{A}=0$,
then (\ref{eq16}) implies that the right side of (\ref{eq15}) is
non-positive. Indeed, introducing the notation $|v|^2=v_{A}v^{A}$,
we have
\begin{equation}
\label{eq17}
4\pi_{A}\psi^{nA}=|\pi_{A} + \psi_{nA}|^2 - |\pi_{A} - \psi_{nA}|^2
\le (\|\alpha\|^2-1) |w^{-}|^2 \le 0.
\end{equation}

Let the differential operator $L$ be defined by Eq.~(\ref{eq14}),
i.e.,
\begin{equation*}
L(\pi_{i},\psi_{ji}) = (\partial_{t}\pi_{i}-\partial^{j}\psi_{ji},
\partial_{t}\psi_{ji}-2\partial_{[j}\pi_{i]}).
\end{equation*}
Let
\begin{align*}
\HH_{L}&=\{ (\pi_{i},\psi_{ji})\in L_{2}(\Omega\times[0,T])\ |\
L(\pi_{i},\psi_{ji})\in L_{2}(\Omega\times[0,T]),\quad \\
&\| (\pi_{i},\psi_{ji}) \|_{\HH_{L}} =
 \| (\pi_{i},\psi_{ji}) \|_{L_2(\Omega\times[0,T])}+
 \| L(\pi_{i},\psi_{ji}) \|_{L_2(\Omega\times[0,T])}\},
\end{align*}
where $ \|(\pi_{i},\psi_{ji})\|_{L_2(\Omega\times[0,T])}=
\| \pi_{i}\|_{L_2(\Omega\times[0,T])}+ \| \psi_{ji}\|_{L_2(\Omega\times[0,T])}$.
It was shown in \cite{R85} (see \cite{R85}, Theorem~2) that the
integration by parts formula holds for $(\pi_{i},\psi_{ji})\in
\HH_{L}$, which implies that the boundary integral in (\ref{eq15}) is
well-defined. Moreover, one can apply the result of \cite{R85}, to
formulate the following
\begin{theorem}
\label{thm1} A unique solution $(\pi_{i}, \psi_{ji})\in \HH_{L}$ exists satisfying (\ref{eq14}),
(\ref{eq14a}) and (\ref{eq16}) (with $g_{A}=0$). In addition, $(\pi_{i},
\psi_{ji})\in C((0,T): L_{2}(\Omega))$ and satisfies the estimate
\begin{equation*}
\sup_{0\le t\le T} [2\| \pi(t)\|_{L_2(\Omega)}+\|\psi(t)\|_{L_2(\Omega)}] \le [2\|
\pi(0)\|_{L_2(\Omega)}+\|\psi(0)\|_{L_2(\Omega)}].
\end{equation*}
\end{theorem}
\begin{remark}
The original theorem in \cite{R85} uses the assumption that the
spatial boundary is $C^1$ while here $\partial \Omega$ is only
piece-wise smooth. However, in \cite{R85} it is suggested how to
generalize the result to a boundary with corners. In general, both
initial and boundary data have to be compatible at the corners to
maintain smoothness of the solution.
\end{remark}
\begin{remark}
A similar statement can be proved using methods of \cite{J91,LP60,LP73,M75,MO75}.
\end{remark}

In the case of inhomogeneous boundary data ($g_{A}(x,t)\neq 0$) we
will assume as in \cite{R85} that the problem is reducible to
the homogeneous case by subtracting an $H^{1}$ function from the
solution. For that, we require that the function $g_{A}(x,t)\in
H^{1/2}(\partial\Omega\times[0,T])$ and satisfies the necessary
compatibility conditions at corners so as to ensure that a vector
field $\hat{g}_{i}\in H^{1}(\Omega\times[0,T])$ exists such that
$\hat{g}_{A}|_{\partial \Omega} = g_{A}/2$. Moreover, we assume
that a vector field $\check{g}_{i}\in H^{1}(\Omega\times[0,T])$
exists such that (a) $\check{g}^A\hat{g}_{A}|_{\partial
\Omega}=0$; (b) at each point of boundary,
$\|\check{g}_{A}\|=\|g_{A}\|/2$ (here $\|\, \cdot \,\|$ is
referencing to the length of the two-vector). From these
two assumptions, it is not difficult to deduce that
\begin{equation*}
n^{j}\varepsilon_{jlA}\check{g}^{l}=(\pm)\hat{g}_{A}=(\pm)g_{A}/2,\qquad
\mbox{on}\ \partial \Omega.
\end{equation*}
We will assume that coefficients of $\check{g}_{i}$ are chosen in
such a way as to have
$n^{j}\varepsilon_{jlA}\check{g}^{l}=\hat{g}_{A}=g_{A}/2$.

It is straightforward to verify that if fields $\hat{g}_{i}$ and
$\check{g}_{i}$ with the above properties exist, the problem
(\ref{eq14}), (\ref{eq14a}), and (\ref{eq16}) can be reduced to
the homogeneous case (with a non-trivial
forcing term in the evolution equations) by introducing the variables
$\tilde{\pi}_{i}=\pi_{i}-\hat{g}_{i}$,
$\tilde{\psi}_{ji}=\psi_{ji}-\varepsilon_{jli}\check{g}^{l}$.

\begin{remark}
In Sections~7 and~8, methods are presented to study properties of
the solution in the case of inhomogeneous data $g_{A}$ in
(\ref{eq16}) not invoking the above reduction. Instead, the
techniques of the Laplace-Fourier transform are used for the
half-plane reduction of the problem \cite{GKO95,KL89,KW06,R06}.
\end{remark}

The result in \cite{R85} implies that boundary integral in
(\ref{eq15}) is defined. However, in (\ref{eq16}) holds, we can
strengthen the result slightly by showing that, for
$(\pi_{i},\psi_{ji})\in \HH_{L}$, the traces
$w^{+}_{A}=\pi_{A}+\psi_{nA}$ and $w^{-}_{A}=\pi_{A}-\psi_{nA}$
are $L_{2}(\partial\Omega)$.
\begin{lemma}
\label{lem4} Let a vector field $\pi_{i}$, and an anti-symmetric
matrix field $\psi_{ji}$ be such that $\pi_{i}\in L_{2}(\Omega)$,
$\partial_{[j}\pi_{i]} \in L_{2}(\Omega)$, and similarly,
$\psi_{ji}\in L_{2}(\Omega)$, $\partial^{j}\psi_{ji} \in
L_{2}(\Omega)$. If, in addition, $\pi_{i}$ and $\psi_{ji}$ satisfy
(\ref{eq16}) with $g_{A}=0$, then
\begin{align*}
\|\pi_{A}-\psi_{nA} \|^2_{L_2(\partial \Omega)}
&\le
\frac{1}{1-\|\alpha\|^2} [\|\pi \|^2_{L_2(\Omega)}+ \| \partial_{[j}\pi_{i]} \|^2_{L_2(\Omega)}+
\|\psi \|^2_{L_2(\Omega)}+ \| \partial^{j}\psi_{ji} \|^2_{L_2(\Omega)}] \\
&:= \frac{1}{1-\|\alpha\|^2}[\|\pi \|_{H^{1}(\curl;\Omega)}+\|\psi \|_{H^{1}(\div;\Omega)}].
\end{align*}
\end{lemma}
\begin{proof}
The following identity can be verified by integration by parts:
\begin{equation*}
\int_{\Omega} \partial_{[j}\pi_{i]}\psi^{ji} dx+\int_{\Omega}
\partial^{j}\psi_{ji}\pi^{i}dx =\int_{\partial\Omega}
\pi_{A}\psi^{nA} d \sigma.
\end{equation*}
Recalling (\ref{eq17}),
\begin{equation*}
\Bigl|\int_{\partial\Omega} \pi_{A}\psi^{nA} d \sigma \Bigr| \ge (1-\|\alpha\|^2)\|\pi_{A}-\psi_{nA} \|^2_{L_2(\partial
\Omega)}.
\end{equation*}
The statement of the lemma follows by applying the Schwartz
inequality to the left side of the integral identity.
\end{proof}

Adjusting notations of \cite{R85} to the variables of system
(\ref{eq14}), one can formulate the following theorem
\begin{theorem}
\label{thm2} Let conditions (\ref{eq16}) be imposed on the
boundary $\partial \Omega\times(0,T)$ and let the vector function
$g_{A}(x,t)\in H^{1/2}(\partial\Omega\times [0,T]))$ be such that
the fields $\hat{g}_{i}$ and $\check{g}_{i}$ exist with the above
properties. Then there exists a unique solution $(\pi_{i},
\psi_{ji})\in \HH_{L}$ to (\ref{eq14}), (\ref{eq14a}). In
addition, $(\pi_{i}, \psi_{ji})\in C((0,T): L_{2}(\Omega))$ and
satisfies the estimate
\begin{equation*}
\sup_{0\le t\le T} [2\| \pi(t)\|_{L_2(\Omega)}+\|\psi(t)\|_{L_2(\Omega)}] \le c\int_{0}^{T}
\|g_{A}\|_{H^{1/2}(\partial\Omega)} dt + [2\|\pi(0)\|_{L_2(\Omega)}+\|\psi(0)\|_{L_2(\Omega)}],
\end{equation*}
where the constant $c$ is independent of $\pi_{i}$, $\psi_{ji}$.
\end{theorem}
\begin{remark}
A similar statement for (\ref{eq14}), (\ref{eq14a}), (\ref{eq16}) can be found in \cite{MO75}.
\end{remark}

\section{Constraint-preserving boundary conditions for the second order equations}

When written in terms of the original variables of
Eq.~(\ref{eq13}), the characteristic speeds and variables read
\begin{align*}
    0:\quad     & \partial_{t}u_{n},\quad \partial_{[A}u_{B]},\\
\pm 1:\quad     & w_{A}^{\pm}=\partial_{t}u_{A} \pm 2\partial_{[n}u_{A]}.
\end{align*}
The maximally dissipative boundary conditions (\ref{eq16})
take the form
\begin{equation}
\label{eq18}
(\partial_{t}u_{A} + 2\partial_{[n}u_{A]}) = \alpha^{B}_{A} (\partial_{t}u_{B} -
2\partial_{[n}u_{B]})+g_{A}.
\end{equation}
Using Theorem~\ref{thm2} one can formulate the following
\begin{theorem}
\label{thm3a} There exists a unique solution to the problem
(\ref{eq13}), (\ref{eq18}). Moreover, the solution obeys the
energy estimate:
\begin{align}
\label{eq18a}
\sup_{0\le t\le T} [&\|\partial_{t}u_{i}\|_{L_2(\Omega)}+2\|\partial_{[j}u_{i]}\|_{L_2(\Omega)}] \nonumber \\
 & \le c\int_{0}^{T}
\|g_{A}\|_{H^{1/2}(\partial\Omega)} dt + [\|
\partial_{t}u_{i}(0)\|_{L_2(\Omega)}+2\|\partial_{[j}u_{i]}(0)\|_{L_2(\Omega)}].
\end{align}
\end{theorem}

We use well-posedness of (\ref{eq13}), (\ref{eq18}) to establish the following
result.
\begin{theorem}
\label{thm3} The constrained evolution problem (\ref{eq1}),
(\ref{eq2}), provided with boundary conditions (\ref{eq18}) is
well-posed, i.e., for any initial data compatible
with the constraint (in the sense of Corollary~\ref{cor1}), a unique solution to
(\ref{eq1}), (\ref{eq2}), (\ref{eq18}) exists and satisfies the estimate of Theorem~\ref{thm3a}.
\end{theorem}
\begin{proof} To construct the solution of (\ref{eq1}) we explore the
relationship between the two evolution Eqs., (\ref{eq1}) and
(\ref{eq13}). Specifically, we use the fact that any solution to
(\ref{eq1}) that satisfies (\ref{eq2}) solves (\ref{eq13}) as
well, and that conversely, any solution to (\ref{eq13}) that
satisfies constraint (\ref{eq2}) solves (\ref{eq1}).

According to Lemma~{\ref{lem2}} the evolution of the constraint
quantity under (\ref{eq13}) is static. This implies that any
solution to (\ref{eq13}) corresponding to constraint-compatible
data automatically satisfies (\ref{eq2}) for all $t$. Thus any
solution to (\ref{eq13}), (\ref{eq18}) obtained from
constraint-compatible initial data solves (\ref{eq1}), (\ref{eq2})
as well. Existence is established. Similarly, multiple solutions
to (\ref{eq1}), (\ref{eq2}), (\ref{eq18}) imply multiple solution
solutions to (\ref{eq13}), (\ref{eq18}) which is impossible in
view of Theorem~\ref{thm3a}. Thus uniqueness is established.
\end{proof}

\begin{remark}
Theorem~\ref{thm3} states that the constrained evolution
problem (\ref{eq1}), (\ref{eq2}) is well-posed if the initial data
is constraint-compatible and two boundary conditions are given in
the form of (\ref{eq18}) (with freely specifiable $g_{A}$).
\end{remark}

It is often advantageous from the numerical point of view to solve
the free evolution problem versus the fully constrained problem.
We recall that only the evolution Eq.~(\ref{eq1}) is solved in the
free evolution problem, while constraint (\ref{eq2}) is monitored
but not actively enforced on the solution. As is discussed in
Section~2, to guarantee that (\ref{eq2}) is satisfied for all
times, the boundary data on variables of (\ref{eq1}) must be given
consistently with Eq.~(\ref{eq9}) that describes evolution of the
constraint quantity. The following theorem gives an example of
radiation-controlling boundary conditions that guarantee
preservation of the constraint.
\begin{theorem}
\label{thm4} Let $\alpha\in \R$, $|\alpha |\le 1$, and the fields
$g$ and $g_{A}$ defined on the boundary be compatible at corners
and satisfy
\begin{equation}
\label{eq19a}
\partial_{t}g=-\partial^{A}g_{A},\quad
g(0)=(1+\alpha)\partial_{t}u_{n}(0) +
(1-\alpha)\partial_{n}u_{n}(0),
\quad \mbox{on}\quad \partial \Omega.
\end{equation}
Then there exists a unique solution to the free evolution problem
(\ref{eq1}) satisfying the constraint compatible initial data (in
the sense of Corollary~\ref{cor1}) and the boundary conditions
\begin{gather}
(\partial_{t}u_{A} + 2\partial_{[n}u_{A]}) = \alpha (\partial_{t}u_{A} -
2\partial_{[n}u_{A]})+g_{A},\nonumber \\
\label{eq19}
(\partial_{t}u_{n} + \partial_{n}u_{n})=-\alpha(\partial_{t}u_{n} -
\partial_{n}u_{n})+g.
\end{gather}
Moreover, the solution satisfies the constraint Eq.~(\ref{eq2}).
\end{theorem}
\begin{proof} To prove existence we recall that by Theorem~{\ref{thm3}}
there exists a unique solution satisfying (\ref{eq1}), (\ref{eq2}), and the first
condition of (\ref{eq19}).

Next we expand $C=\partial^{i}u_{i}=\partial_{n}u_{n}+\partial^{A}u_{A}$ using
normal and tangential derivatives and consider
\begin{equation}
\label{eq20}
\partial_{t}C+\partial_{n}C=\partial_{t}\partial_{n}u_{n}+
\partial_{t}\partial^{A}u_{A}+\partial_{n}\partial_{n}u_{n}+
\partial_{n}\partial^{A}u_{A}.
\end{equation}
Taking the normal component of (\ref{eq1}) we have
\begin{equation*}
\partial^{2}_{t}u_{n}=\partial_{n}\partial_{n}u_{n}+\partial^{A}\partial_{A}u_{n}.
\end{equation*}
Solving the last expression for $\partial_{n}\partial_{n}u_{n}$
and substituting into (\ref{eq20}), we obtain
\begin{equation*}
\partial_{t}C+\partial_{n}C=\partial_{t}\partial_{n}u_{n}+\partial_{t}\partial^{A}u_{A}
+\partial_{t}\partial_{t}u_{n}-\partial^{A}\partial_{A}u_{n}+
\partial_{n}\partial^{A}u_{A}.
\end{equation*}
By commuting derivatives and rearranging terms in the last
expression, we have
\begin{equation}
\label{eq20a}
\partial_{t}C+\partial_{n}C=\partial_{t}(\partial_{t}u_{n}+\partial_{n}u_{n})+
\partial^{A}(\partial_{t}u_{A}+2\partial_{[n}u_{A]}).
\end{equation}

Similarly,
\begin{equation}
\label{eq20b}
\partial_{t}C-\partial_{n}C=-\partial_{t}(\partial_{t}u_{n}-\partial_{n}u_{n})+
\partial^{A}(\partial_{t}u_{A}-2\partial_{[n}u_{A]}).
\end{equation}

Let us show that the second condition of (\ref{eq19}) is a
consequence of (\ref{eq1}), (\ref{eq2}), and the first condition.
First of all since (\ref{eq2}) is satisfied, both
$\partial_{t}C+\partial_{n}C=0$ and
$\partial_{t}C-\partial_{n}C=0$. Multiplying (\ref{eq20b}) by
$\alpha$ and subtracting it from (\ref{eq20a}), we obtain
\begin{equation*}
0=\partial_{t}[(\partial_{t}u_{n}+\partial_{n}u_{n})+\alpha(\partial_{t} u_{n}-\partial_{n} u_{n})] +
\partial^{A}[ (\partial_{t}u_{A}+2\partial_{[n}u_{A]})-\alpha
(\partial_{t}u_{A}-2\partial_{[n}u_{A]})].
\end{equation*}
In view of the first condition of (\ref{eq19}) the latter can be
rewritten as
\begin{equation*}
0=\partial_{t}[(\partial_{t}u_{n}+\partial_{n}u_{n})+\alpha(\partial_{t} u_{n}-\partial_{n} u_{n})] +
\partial^{A}g_{A}.
\end{equation*}
Using (\ref{eq19a}) we have (\ref{eq19}) by integration in time. Existence is proved.

To prove uniqueness, it is sufficient to show that any solution to
(\ref{eq1}), (\ref{eq19}) preserves the constraint, so that we can
employ the second part of the proof of Theorem~{\ref{thm3}}.

By multiplying (\ref{eq9}) with $\partial_{t}C$, taking the
integral over $\Omega$, and integrating by parts,
\begin{equation*}
\int_{\Omega}(\partial^2_{t}C )(\partial_{t}C )dx=
-\int_{\Omega}(\partial_{i}C)(\partial^{i}\partial_{t}C) dx+
\int_{\partial \Omega} (\partial_{n} C)(\partial_{t}C)d\sigma,
\end{equation*}
where $n_{i}$ is the outward pointing unit normal vector to the
boundary. By commuting derivatives and reorganizing terms we
rewrite the last equation as
\begin{equation}
\label{eq10a}
\partial_{t}[\|\partial_{t}C\|^2_{L_{2}(\R^3)}+\|\partial_{i}C\|^2_{L_{2}(\R^3)}]
=\frac{1}{2}\int_{\partial \Omega}[(\partial_{t}C+\partial_{n}C)^{2}-(\partial_{t}C-\partial_{n}C)^2]d\sigma.
\end{equation}
An argument similar to that of Corollary~\ref{cor1} can be
used to show that if $C$ satisfies
\begin{equation}
\label{eq10b}
(\partial_{t}C+\partial_{n} C)=\alpha (\partial_{t}C-\partial_{n} C)\quad \mbox{on}\ \partial \Omega,
\end{equation}
with some constant $|\alpha|\le 1$, the energy quantity in (\ref{eq10a}) is not
increasing. Thus the constraint quantity remains zero if it is zero initially.

Indeed, (\ref{eq10b}) follows from (\ref{eq20a}) and (\ref{eq20b}) by (\ref{eq19}) and
(\ref{eq19a}). Uniqueness is proved.
\end{proof}

\begin{remark}
To establish Theorem~{\ref{thm4}} we use the assumption of the
exact compatibility of data. Such assumption, however, is not
practical in numerical applications where the equations are
satisfied only approximately, and where one has to worry about
propagation of small constraint violations. The next result
generalizes Theorem~\ref{thm4} to the case of arbitrary data.
\end{remark}

We rewrite Eq.~(\ref{eq1}) as
\begin{equation}
\label{eq21a1}
\partial^{2}_{t}u_{i}=2\partial^{l}\partial_{[l}u_{i]}+\partial_{i}\partial^{l}u_{l}.
\end{equation}
We introduce the new variables $\pi_{i}=\partial_{t}u_{i}$,
$\psi_{ji}=2\partial_{[j}u_{i]}$ and $\varphi=\partial^{l}u_{l}$,
and decompose (\ref{eq21a1}) into
\begin{align}
\partial_{t}u_{i}&=\pi_{i},\nonumber \\
\partial_{t}\pi_{i}&=\partial^{j}\psi_{ji}+\partial_{i}\varphi,\nonumber \\
\partial_{t}\psi_{ji}&=2\partial_{[j}\pi_{i]},\nonumber\\
\label{eq21b}
\partial_{t}\varphi&=\partial^{l}\pi_{l}.
\end{align}
The resulting first order system is symmetric hyperbolic. Its
characteristic speeds and fields in the direction $n_{i}$ are
given by
\begin{align*}
    0:\quad     & \psi_{AB},\\
\pm 1:\quad     & w_{A}^{\pm}=\pi_{A} \pm \psi_{nA},\quad w^{\pm}=\pi_{n} \pm \varphi.
\end{align*}
Notice that the first two boundary conditions in (\ref{eq19})
remain the same for (\ref{eq21b}) and, in fact, take the familiar
form of (\ref{eq16}) with the substitution
$\alpha_{ij}=\alpha\delta_{ij}$. The last condition, however, can
not be implemented directly since $\partial_{n}u_{n}$ is not among
the variables $\psi_{ij}$ and $\varphi$ (but is closely related to
$\varphi$). To formulate an equivalent condition by considering Eqs.~(\ref{eq21b}) at the face of the
boundary we derive the following identity:
\begin{equation}
\label{eq21c}
(\partial_{t}\varphi \pm \partial_{n}\varphi) =
\pm(\partial_{t}
\pi_{n}\pm\partial_{n}\pi_{n})+\partial^{A}(\pi_{A}\pm\psi_{nA}).
\end{equation}

Let the differential operator $\tilde{L}$ be defined by Eqs.~(\ref{eq21b}), i.e.,
\begin{equation*}
\tilde{L}(\pi_{i},\psi_{ji},\varphi)=(\partial_{t}\pi_{i}-\partial^{j}\psi_{ji}-\partial_{i}\varphi,
\partial_{t}\psi_{ji}-2\partial_{[j}\pi_{i]},\partial_{t}\varphi-\partial^{l}\pi_{l}).
\end{equation*}
As in Section~4, we define
\begin{align*}
\HH_{\tilde{L}}&=\{ (\pi_{i},\psi_{ji},\varphi)\in L_{2}(\Omega\times[0,T])\ |\
\tilde{L}(\pi_{i},\psi_{ji},\varphi)\in L_{2}(\Omega\times[0,T]),\quad \\
&\| (\pi_{i},\psi_{ji},\varphi) \|_{\HH_{\tilde{L}}} =
 \| (\pi_{i},\psi_{ji},\varphi) \|_{L_2(\Omega\times[0,T])}+
 \| L(\pi_{i},\psi_{ji},\varphi) \|_{L_2(\Omega\times[0,T])}\}.
\end{align*}

\begin{theorem}
\label{thm4b} Let $\alpha\in \R$, $|\alpha |\le 1$, and the fields
$g$ and $g_{A}$ be defined on the boundary and compatible at
corners and assume that $g_{A}$ meets the assumptions of
Theorem~{\ref{thm2}}. Let, in addition, functions $g$ and $g_{A}$
be such that $(\partial_{t}g + \partial^{A}g_{A})\in
H^{1/2}(\partial \Omega)$. Then there exists a unique solution
$(\pi_{i}, \psi_{ji}, \varphi)\in \HH_{\tilde{L}}$ to
(\ref{eq21b}) satisfying the boundary conditions
\begin{gather}
\label{eq21d}
w^{+}_{A} = \alpha w^{-}_{A}+g_{A},\nonumber \\
\label{eq21d1}
(\partial_{t}\varphi + \partial_{n}\varphi) = \alpha (\partial_{t}\varphi - \partial_{n}\varphi)
- (\partial_{t}g + \partial^{A}g_{A}).
\end{gather}
Moreover, the solution satisfies the energy estimate
\begin{align}
\sup_{0\le t\le T}& [2\| \pi(t)\|_{L_2(\Omega)}+\|\psi(t)\|_{L_2(\Omega)}+\| \varphi(t) \|_{L_{2}(\Omega)}] \nonumber \\
 &\le c \int_{0}^{T} \|g_{A}\|_{H^{1/2}(\partial\Omega)} dt
   + c_2\e^T(c_1 \int_{0}^{T} \|\partial_{t}g + \partial^{A}g_{A} \|_{H^{1/2}(\partial\Omega)} dt \nonumber \\
 &{} + [\| \partial^{l}\partial_{t}u_{l}(0) \|_{L_2(\Omega)} + \| \partial^{l}u_{l}(0) \|_{L_2(\Omega)}]) \nonumber \\
 \label{eq21g}
 &{} + [2\|\pi(0)\|_{L_2(\Omega)}+\|\psi(0)\|_{L_2(\Omega)}],
\end{align}
where the constants $c$, $c_1$, $c_2$ are independent of $\pi_{i}$, $\psi_{ji}$, $\varphi$.
\end{theorem}
\begin{proof}
By differentiating the last of Eqs.~(\ref{eq21b}) with respect to time and
substituting the second in the result, we obtain
\begin{equation}
\label{eq21e}
\partial^{2}_{t}\varphi=\partial^{l}\partial_{l}\varphi.
\end{equation}
The initial data for (\ref{eq21e}) is given by
$\varphi(0)=\partial^{l}u_{l}(0)$ and
$\partial_{t}\varphi(0)=\partial^{l}\partial_{t}u_{l}(0)$. By
employing the first order reduction, it is straightforward to
establish existence and uniqueness of the solution. Indeed, by
introducing $\partial_{t}\varphi=\xi$ and
$\zeta_{l}=\partial_{l}\varphi$, we obtain the first order
symmetric hyperbolic decomposition of (\ref{eq21e}). Notice that,
when written in terms of $\xi$ and $\zeta_{l}$, (\ref{eq21d1})
represents the maximally nonnegative boundary conditions
(which are in general non-homogeneous). To establish the existence, it is
sufficient to notice that the assumption $(\partial_{t}g +
\partial^{A}g_{A})\in H^{1/2}(\partial\Omega)$ guarantees the existence
of functions $\hat{h}$ and $\check{h}_{l}$ such that
$\tilde{\xi}=\xi-\hat{h}$,
$\tilde{\zeta}_{l}=\zeta_{l}-\check{h}_{l}$ satisfy the
homogeneous conditions. Indeed, one could take $\hat{h}$ to be the
$H^{1}$ extension of $(\partial_{t}g +
\partial^{A}g_{A})/2$, and set $\check{h}_{l}=\partial_{l}\rho$, where
$\rho$ solves $\partial^{l}\partial_{l}\rho=0$, $(\partial/\partial n) \rho|_{\partial \Omega} = (\partial_{t}g +
\partial^{A}g_{A})/2$. By completing the argument as in Section~4, we
arrive at the estimate
\begin{align}
\sup_{0\le t\le T} [\| \partial_{t}\varphi(t)\|_{L_2(\Omega)}&+\|\partial_{l}\varphi(t)\|_{L_2(\Omega)}] \le
 c_1 \int_{0}^{T} \|\partial_{t}g + \partial^{A}g_{A} \|_{H^{1/2}(\partial\Omega)} dt \nonumber \\
\label{eq21f}
&{}+ [\| \partial^{l}\partial_{t}u_{l}(0)\|_{L_2(\Omega)}+\|\partial^{l}u_{l}(0)\|_{L_2(\Omega)}].
\end{align}
Finally, by repeating the constructions of Theorems~\ref{thm2}
and~\ref{thm1}, with the only difference being that
$\partial_{i}\varphi$ is now the forcing term in the second equation
of (\ref{eq21b}), we establish the existence result and the
estimate (\ref{eq21g}). The term $\| \varphi(t)
\|_{L_{2}(\Omega)}$ is added to (\ref{eq21g}) by considering the inequality
\begin{equation*}
\partial_{t}\|\varphi\|^2_{L_2(\Omega)} = 2(\partial_{t}\varphi, \varphi) \le \|\partial_{t}\varphi\|^2 + \| \varphi \|^2
\end{equation*}
and using (\ref{eq21f}).
\end{proof}
\begin{remark}
It seems to be more natural to seek a solution to the
problem (\ref{eq21b}) in the space $\KK_{\tilde{L}}=\{ (\pi_{i},
\psi_{ji}, \varphi )\in L_{2}(\Omega)\ | \ \tilde{L}(\pi_{i},
\psi_{ji},\varphi)\in H^{-1}(\Omega)\}$ (see \cite{R85}). This is
possible if one extends the results of \cite{R85} to weaker spaces
and constructs the solution $\varphi\in L_{2}(\Omega)$,
$\partial^{l}\partial_{l}\varphi\in H^{-2}(\Omega)$ to
(\ref{eq21e}). A solution in this form does not require the
``extra'' smoothness assumption $(\partial_{t}g +
\partial^{A}g_{A})\in H^{1/2}(\partial \Omega)$. Instead, it requires the
milder assumption $(\partial_{t}g +
\partial^{A}g_{A})\in H^{-1/2}(\partial \Omega)$ which is the consequence
of the other assumptions of Theorem~\ref{thm4b}. It seems likely
that one could use the general scheme of \cite{R85} to prove the
existence without the extra smoothness assumption. The proof, however, is
rather technically involved. We, therefore, do not attempt it in
this work.
\end{remark}

Theorem~\ref{thm4b} is used to formulate a well-posedness result
for the free evolution problem (\ref{eq1}), (\ref{eq19}).
\begin{theorem}
\label{thm4c} Let $\alpha\in \R$, $|\alpha |\le 1$, and the fields
$g$ and $g_{A}$ satisfy the assumptions of Theorem~{\ref{thm4b}}.
Let, in addition, the initial and boundary data be compatible, i.e.,
\begin{equation}
\label{eq21i}
g(0)=(1+\alpha)\partial_{t}u_{n}(0) +
(1-\alpha)\partial_{n}u_{n}(0),
\quad \mbox{on}\quad \partial \Omega.
\end{equation}
Then a unique solution $u_{i}\in L_{2}(\Omega)$,
$\partial_{[j}u_{j]}\in L_{2}(\Omega)$, $\partial^{l}u_{l}\in
L_{2}(\Omega)$ exists to (\ref{eq21a1}) satisfying boundary
conditions (\ref{eq19}). Moreover, the solution satisfies the
estimate
\begin{align*}
\sup_{0\le t\le T}& [2\| \partial_{t}u_{i}\|_{L_2(\Omega)}+\|\partial_{[j}u_{i]}\|_{L_2(\Omega)}+
 \| \partial^{l}u_{l} \|_{L_{2}(\Omega)}] \nonumber \\
 &\le c \int_{0}^{T} \|g_{A}\|_{H^{1/2}(\partial\Omega)} dt
   + c_2\e^T(c_1 \int_{0}^{T} \|\partial_{t}g + \partial^{A}g_{A} \|_{H^{1/2}(\partial\Omega)} dt \nonumber \\
 &{} + [\| \partial^{l}\partial_{t}u_{l}(0) \|_{L_2(\Omega)} + \| \partial^{l}u_{l}(0) \|_{L_2(\Omega)}])
  +
  [2\|\partial_{t}u_{i}(0)\|_{L_2(\Omega)}+\|\partial_{[j}u_{i]}(0)\|_{L_2(\Omega)}].
\end{align*}
\end{theorem}
\begin{proof}
Existence of the solution follows from the first order reduction
and Theorem~\ref{thm4b}. It should be noted, however, that the
first order reduction (\ref{eq21b}) involves the constraint
$C_{ji} = \psi_{ji} - 2\partial_{[j} u_{i]} = 0$. Thus a solution
to (\ref{eq21b}) only yields a solution to (\ref{eq21a1}) if this
constraint is satisfied. However, this poses no problems since
(\ref{eq21b}) yields that $\partial_t C_{ji} = 0$. Since $C_{ji}
= 0$ for the initial data $u_{i}(0)$, $\psi_{ji}(0)$, the latter
implies that $C_{ji} = 0$ is automatically
satisfied for future times. Thus the component $u_{i}$ of the
solution to (\ref{eq21b}) gives a solution to (\ref{eq21a1}).
Finally, one has to check that second part of (\ref{eq19}) is
satisfied by $u_{i}$. In view of (\ref{eq21i}), the latter follows
from identities (\ref{eq21c}) and conditions (\ref{eq21d1}) by
integration in time. For the uniqueness proof, it is sufficient to
notice that any solution to (\ref{eq21a1}), (\ref{eq19}) solves
(\ref{eq21b}) in the sense of the first order reduction described
above. Moreover, the variable $\varphi$ satisfies (\ref{eq21d1})
in view of identities (\ref{eq21c}). Then the uniqueness follows from
Theorem~\ref{thm4b}.
\end{proof}

\section{The full gradient estimate for the static constraint evolution system}

Energy estimates of the type (\ref{eq18a}) are often considered to
be a disadvantage for the numerical solution of (\ref{eq1}),
because they do not control all components of the gradient. The
main objection to (\ref{eq18a}) comes from the fact that if lower
order (nonlinear) terms are present in the equations that include
components of the gradient not controlled by (\ref{eq18a}),
coupling to such not-controlled terms may threaten the stability of
the entire system.

Next we will show how energy estimate (\ref{eq18a}) can be improved
by employing the constraint preservation property. It has to be
emphasized, however, that the existence and uniqueness
results above suggest that energy estimate (\ref{eq18a}) is
accurate for the problem and may not be improved unless one
further restricts the class of solutions by specifying special boundary or
initial data, or making additional assumptions about the equations
(e.g., the gauge choice of \cite{RS05} for the fat Maxwell's
system).

The following energy argument can be applied to a static
constraint evolution system to provide additional energy
estimates. Rewrite (\ref{eq13}) in the divergence form,
\begin{equation*}
\partial^{2}_{t}u_{i}=\partial^{j}(\partial_{j}u_{i}-\delta_{ji}\partial^{p}u_{p}).
\end{equation*}
Contracting the last expression with $\partial_{t}u^{i}$ and
integrating by parts, we have
\begin{equation*}
\frac{1}{2}\partial_{t}\|\partial_{t} u\|^2_{L_2(\Omega)}=
-\int_{\Omega}(\partial_{j}u_{i}-\delta_{ji}\partial^{p}u_{p})(\partial^{j}\partial_{t}u^{i})dx
+\int_{\partial\Omega} (\partial_{n} u_{i} -
n_{i}\partial^{p}u_{p})(\partial_{t}u^{i})d\sigma.
\end{equation*}
Recalling that $C=\partial^{p}u_{p}$ we expand the identity by
adding and subtracting terms in the first integral:
\begin{align*}
\frac{1}{2}\partial_{t}\|\partial_{t} u \|^2_{L_2(\Omega)}&=
-\int_{\Omega}(\partial_{j}u_{i}-\delta_{ji}C)(\partial^{j}\partial_{t}u^{i}-\delta^{ji}\partial_{t}C)dx\\
&{} -\int_{\Omega}(\partial_{j}u_{i}-\delta_{ji}C)(\delta^{ji}\partial_{t}C)dx
+\int_{\partial\Omega} (\partial_{n} u_{i} -
n_{i}\partial^{p}u_{p})(\partial_{t}u^{i})d\sigma.
\end{align*}
The latter becomes after rearrangement
\begin{align}
\frac{1}{2}\partial_{t}[\|\partial_{t} u\|^{2}_{L_2(\Omega)} +
\| \partial_{j}u_{i}-\delta_{ji}C \|^2_{L_2(\Omega)}]
&=-\int_{\Omega}(\partial_{j}u_{i}-\delta_{ji}C)(\delta^{ji}\partial_{t}C)dx \nonumber \\
\label{eqX1}
&{}+\int_{\partial\Omega} (\partial_{n} u_{i} -
n_{i}\partial^{p}u_{p})(\partial_{t}u^{i})d\sigma.
\end{align}
Let us for a moment assume that a solution to Eq.~(\ref{eq13}) is
defined in the entire $\R^3$ and is integrable. Then the boundary
integral in the last expression drops out. Also, if the solution
$u_{i}$ is obtained from the constraint compatible initial data, then
the first integral at the right side as well as all other terms in
$C$ drop out due to the trivial constraint preservation (see
Lemma~\ref{lem2}). As a result, we obtain an estimate on all
components of the solution's gradient.

If the initial data is not constraint compatible, e.g., is
perturbed by roundoff error, and one is dealing with a bounded
domain $\Omega\subset \R^3$ one can rewrite (\ref{eqX1}) as
\begin{equation*}
\partial_{t}[\|\partial_{t} u_{i}\|^{2}_{L_2(\Omega)} +
\| \partial_{j}u_{i}-\delta_{ji}C \|^2_{L_2(\Omega)} -
2\|C\|^2_{L_2(\Omega)}] = 2\int_{\partial\Omega} (\partial_{n} u_{i} -
n_{i}\partial^{p}u_{p})(\partial_{t}u^{i})d\sigma.
\end{equation*}
Due to the static constraint evolution,
$C=\partial_{t}C(0)t+C(0)$. In other words, $\|C\|$ is completely
determined by the initial data. Thus the left side gives the norm
of the gradient of the solution up to some integral terms
depending on the initial values $u_{i}(0)$,
$\partial_{t}u_{i}(0)$. We define $\epsilon = \|\partial_{t}
u_{i}\|^{2}_{L_2(\Omega)} + \| \partial_{j}u_{i}-\delta_{ji}C
\|^2_{L_2(\Omega)} - 2\|C\|^2_{L_2(\Omega)}$ noting that the right
side of the identity contains a boundary integral and is thus
dependent on the particular boundary conditions.

Rewriting the integrand in terms of normal and tangential
components and simplifying, we have
\begin{equation}
\label{eq21}
\partial_{t}\epsilon = 2 \int_{\partial \Omega} (\partial_{n}
u_{A}\partial_{t}u^{A} + \partial^{A}u_{A}\partial_{t}u_{n}).
\end{equation}
It may look as if the boundary integral requires conditions on all
three components of $u_{i}$ but this is not so. In fact, the
energy $\epsilon$ is conserved if we require the solution to satisfy
\begin{equation}
\label{eq21a}
u_{A}\Big|_{\partial\Omega}=0.
\end{equation}
For, the integrand on the right side vanishes. Therefore the
full gradient norm of the solution to (\ref{eq13}), (\ref{eq21a}) grows
at a rate not faster than $t^2$.

Another interesting example is given by the boundary conditions:
\begin{equation}
\label{eq22}
(\partial_{n} u_{A}-\partial_{A}u_{n})|_{\partial\Omega}=0.
\end{equation}
Notice that taking the normal component of (\ref{eq13}) at the
boundary, we obtain
\begin{equation*}
\partial^{2}_{t}u_{n}=\partial_{n}\partial_{n}u_{n}+\partial^{A}\partial_{A}u_{n}.
\end{equation*}
Due to the static constraint evolution,
$\partial^{i}u_{i}=\partial_{t}C(0)t+C(0)$, which implies that
\begin{equation*}
\partial_{n}\partial_{n}u_{n}=-\partial_{n}\partial^{A}u_{A}+\partial_{n}(\partial_{t}C(0)t+C(0)).
\end{equation*}
Combining the last two equations and using (\ref{eq22}),
\begin{equation*}
\partial^{2}_{t} u_{n} = \partial^{A}\partial_{A} u_{n}
                         - \partial_{n}\partial^{A}u_{A} +
                         \partial_{n}(\partial_{t}C(0)t+C(0))=
                         \partial_{n}(\partial_{t}C(0)t+C(0)).
\end{equation*}
In other words, we have showed that the component $u_{n}$ at the
boundary is determined by the initial data. If
the initial data is such that (a) $u_{n}|_{\partial \Omega} =
\partial_{t}u_{n}|_{\partial \Omega}=0$, and (b) the data is
constraint compatible (so $C\equiv 0$), then the last identity
implies that $u_{n}|_{\partial \Omega}=0$. Together with
(\ref{eq22}), this implies that the boundary integral in
(\ref{eq21}) vanishes. (Notice that assumption (a) overlaps with
the well-posedness result of \cite{RS05} for the fat Maxwell's
system.)

\begin{remark}
A similar full gradient estimate can be derived from the
identity which is true for all $H^{1}$ vector functions (not
necessarily solutions of (\ref{eq13})):
\begin{equation*}
\|\partial_{[j}u_{i]}\|^2_{L_2(\Omega)}+\|\partial^{i}u_{i}\|^2_{L_2(\Omega)}
+\int_{\partial \Omega}(\partial_{j}u_{i}n^{i})u^{j}
=\|\partial_{(j}u_{i)}\|^2_{L_2(\Omega)}
+\int_{\partial \Omega}(\partial^{i}u_{i})n^{j}u_{j}.
\end{equation*}
Using this identity, a full gradient estimate can be developed for
the radiative condition (\ref{eq18}). The estimate does not need
(a), but requires a special assumption, namely, the half plane
reduction for the problem. We investigate this further in the next
section.
\end{remark}

\section{Comparison stability analysis by Laplace-Fourier transform}

In this section we make a comparison of the data derived by the
static constraint evolution reduction to the
constraint-preserving boundary conditions obtained by the standard
methods using the subordinate constraint evolution equation (cf.,
\cite{A04,AT06,CPST03,KLS04,GMG04b,LSKPST04,RS05}). In \cite{RS05}
a general class of constraint-preserving boundary conditions is
derived for a symmetric hyperbolic reduction of the fat Maxwell's
system and the well-posedness is proved for a subclass of these
conditions (which, in fact, overlaps with (\ref{eq19}) in two
equations).  We will study (\ref{eq19}) and the conditions of
\cite{RS05} to see if any significant difference can be observed
in the Lapalace-Fourier analysis.

From (\ref{eq10a}) it is apparent that as soon as either of the
three conditions is satisfied,
\begin{equation}
\label{eq10} C=0, \quad \partial_{n}C=0,\quad
\mbox{or} \quad \partial_{t} C + \partial_{n} C=0,\qquad
\mbox{on}\quad \partial \Omega,
\end{equation}
the constraint quantity is preserved.

First, we consider the classical examples of Dirichlet-Neumann
constraint-preserving boundary conditions (e.g.,
\cite{A04,AT06,CPST03,GMG04b,SW03}). Let $g_{A}$ be a sufficiently
smooth compatible vector field tangential to the boundary. It can
be easily verified that the following boundary conditions imply
$C|_{\partial\Omega}=0$:
\begin{equation}
\label{eq10c1}
u_{A}=g_{A},\qquad \partial_{n}u_{n}=-\partial^{A}g_{A}.
\end{equation}
To construct the boundary condition that enforces
$\partial_{n}C=0$, recall that
\begin{equation*}
\partial_{n} C= \partial_{n}(\partial_{n} u_{n}+\partial^{A} u_{A})=0.
\end{equation*}
By trading the second normal derivative of $u_{n}$ from
(\ref{eq1}) for the temporal and tangential derivatives, one
obtains
\begin{equation}
\label{eq10c2}
\partial^2_{t} u_{n} -
\partial^{A}\partial_{A}u_{n}+\partial_{n} \partial^{A} u_{A}=0 \quad \mbox{on} \quad \partial\Omega.
\end{equation}
One can either impose (\ref{eq10c2}) as a (second order)
differential boundary condition, or to evolve it along the
boundary using the initial data and compatibility conditions at
corners to produce the value of $u_{n}$ at the boundary. In this
case, the constraint-preserving boundary condition takes the form
\begin{equation}
\label{eq10c3}
\partial_{n}u_{A} = h_{A}, \qquad
u_{n}=\hat{u}_{n},
\end{equation}
where $h_{A}$ is a given tangential vector field and $\hat{u}_{n}$
is the solution of (\ref{eq10c2}).

Finally, let us construct the examples of radiative conditions similar
to one that can be found in \cite{RS05} (in which the analog of
(\ref{eq10c3}) is also considered). Substituting $C=\partial_{n}
u_{n}+\partial^{A}u_{A}$ into the third condition of (\ref{eq10}),
commuting derivatives, and trading the second normal derivative for
the second temporal derivatives in (\ref{eq1}), we have
\begin{equation}
\label{eq10c4}
\partial_{t}(\partial_{t}u_{n} + \partial_{n}u_{n})
-\partial^{A}\partial_{A}u_{n}+\partial^{A}(\partial_{t}u_{A}+\partial_{n} u_{A})=0.
\end{equation}
As in \cite{RS05} we treat this equation as an evolution equation
for $\partial_{t}u_{n} + \partial_{n} u_{n}$. It follows that the
following conditions are constraint-preserving:
\begin{equation}
\label{eq10c5}
\partial_{t}u_{A}+\partial_{n} u_{A}=f_{A},\quad
\partial_{t}u_{n}+\partial_{n} u_{n}=f;\quad
\mbox{on} \quad \partial \Omega.
\end{equation}
Here $f=\partial_{t}u_{n} + \partial_{n} u_{n}$ is
obtained by solving (\ref{eq10c4}).

Unfortunately, the equation for the forcing term $f$ does not
decouple from the system. We will still include
(\ref{eq10c5}) in the analysis since it can be implemented
numerically.

We now consider the half space $x \in \R^{3}$, $x_{1}\ge 0$ and
write the first order reduction of Eq.~(\ref{eq1}):
\begin{align}
\label{eq30}
\partial_{t}u_{i}&=\pi_{i}; \quad \varphi_{ki}=\partial_{k}u_{i}; \nonumber\\
\partial_{t}\pi_{i}&=\partial^{j}\varphi_{ji}; \nonumber \\
\partial_{t}\varphi_{ji}&=\partial_{j}\pi_{i}.
\end{align}
Boundary conditions (\ref{eq19}), (\ref{eq10c1}), (\ref{eq10c3}),
(\ref{eq10c5}) are imposed at $x_{1}=0$. In terms of the first
order variables these boundary conditions read (here $g_{A}$, $g$,
$h_{A}$, $h$, and $f_{A}$ are free functions)
\begin{gather}
\label{eq31}
\theta\pi_{x} - \varphi_{xx}=g,\quad
\pi_{A} - \theta(\varphi_{xA}-\varphi_{Ax})=g_{A}, \quad
\theta=\frac{1+\alpha}{1-\alpha},\quad |\alpha|<1, \\
\label{eq32}
\varphi_{xx} = h,\qquad \pi_{A} = h_{A},\\
\label{eq33}
\pi_{x}=h, \qquad \varphi_{xA} = h_{A},\\
\label{eq34}
\partial_{t}(\pi_{x} - \varphi_{xx}) - \partial^{A}(f_{A}+\varphi_{Ax})=0,\quad
\pi_{A}-\varphi_{xA}=f_{A}.
\end{gather}
Notice that the right sides $g_{A}$, $g$, $h_{A}$, and $h$ are no
longer required to be compatible with the constraint. This
reflects the situation common in numerical calculations when the
constraint-preserving boundary conditions are perturbed by the
discretization errors.

We seek a solution of (\ref{eq30}) as a superposition of modes
\begin{equation}
\label{eq35}
\left(\begin{array}{c}
\pi_{i} \\
\varphi_{ji}
\end{array}
\right) =
\left(\begin{array}{c}
\tilde{\pi}_{i}(x_{1}) \\
\tilde{\varphi}_{ji}(x_{1})
\end{array}
\right)\exp(st+\imath\omega_{A}x^{A}),
\end{equation}
where $s\in \C$, $\Re(s)>0$, $\omega_{A}$ is a real two-vector, and
fields $\tilde{\pi}_{i}(x_{1})$, $\tilde{\varphi}_{ji}(x_{1})$
are differentiable. By inserting (\ref{eq35}) into the last two
equations of (\ref{eq30}), we obtain the system of ordinary
differential equations:
\begin{align}
s\tilde{\pi}_{i}(x_{1})&=\partial_{x}\tilde{\varphi}_{xi}(x_{1})+
\imath\omega^{A}\tilde{\varphi}_{Ai}(x_{1}); \nonumber \\
s\tilde{\varphi}_{xi}(x_{1})&=\partial_{x}\tilde{\pi}_{i}(x_{1}); \nonumber \\
\label{eq36}
s\tilde{\varphi}_{Ai}(x_{1})&=\imath\omega_{A}\tilde{\pi}_{i}(x_{1}).
\end{align}
Using the fact that $\Re(s)>0$, we solve the last equation of
(\ref{eq36}) for $\tilde{\varphi}_{Ai}$. Substituting the
obtained expression into the first equation, we have
\begin{align}
\partial_{x}\tilde{\pi}_{i}(x_{1})&=
s\tilde{\varphi}_{xi}(x_{1}),\nonumber \\
\label{eq37}
\partial_{x}\tilde{\varphi}_{xi}(x_{1})&=([s^2+|\omega|^{2}]/s)\tilde{\pi}_{i}(x_{1}).
\end{align}
The system's matrix has eigenvalues $\mu^{\pm}=\pm
\sqrt{s^2+|\omega|^2}$ (the branch of the square root is chosen
such that $\Re(\mu^{+})>0$), and eigenvectors $r^{-1}(s v_{i},
\pm\sqrt{s^2+|\omega|^2} v_{i})$. Here $v_{i}$ is a unit
three-vector, $r>0$ is the normalization factor
$r^2=|s^2+|w|^2|+|s|^2$, where $|s|=\sqrt{s\bar{s}}$.

The most general solution of (\ref{eq37}) that is bounded at infinity takes the form
\begin{equation}
\label{eq38}
\left(\begin{array}{c}
\tilde{\pi}_{i}(x_{1}) \\
\tilde{\varphi}_{xi}(x_{1})
\end{array}
\right) =\sum_{\nu=1}^{3}
\sigma_{\nu} w_{\nu} \exp(-\sqrt{s^2+|\omega|^{2}}\, x_{1}),
\end{equation}
where $w_{\nu}=r^{-1}( s (v_{\nu})_{i}, -\sqrt{s^2+|\omega|^2}
(v_{\nu})_{i} )$, and $(v_{1})_{i}=(1,0,0)$,
$(v_{2})_{i}=(0,1,0)$, $(v_{3})_{i}=(0,0,1)$. Notice that in
(\ref{eq38}) only the components corresponding to the eigenvalues
with negative real part are taken.

The coefficients $\sigma_{\nu}$ are to be determined from the
boundary conditions (\ref{eq31})--(\ref{eq34}). By taking the
Laplace transform with respect to time and the Fourier transform
in variables $x_{A}=(x_{2},x_{3})$, and using the last equation of
(\ref{eq36}) to eliminate $\tilde{\varphi}_{Ai}$, we rewrite
boundary conditions (\ref{eq31})--(\ref{eq34}) as
\begin{gather}
\theta\tilde{\pi}_{x}(0) - \tilde{\varphi}_{xx}(0)=\tilde{g}, \
\tilde{\pi}_{A}(0) - \theta(\tilde{\varphi}_{xA}(0)-\frac{\imath \omega_{A}}{s} \tilde{\pi}_{x}(0))=\tilde{g}_{A}, \
\theta=\frac{1+\alpha}{1-\alpha},\  |\alpha|<1,
\nonumber \\
\tilde{\varphi}_{xx}(0) = \tilde{h},\qquad \tilde{\pi}_{A}(0) = \tilde{h}_{A}, \nonumber\\
\tilde{\pi}_{x}(0)= \tilde{h},\qquad \tilde{\varphi}_{xA}(0) = \tilde{h}_{A}, \nonumber\\
\label{eq39}
\frac{s^2+|\omega|^2}{s}\tilde{\pi}_{x}(0) - s\tilde{\varphi}_{xx}(0) = \imath
\omega^{A}\tilde{f}_{A}, \qquad
\tilde{\pi}_{A}(0)-\tilde{\varphi}_{xA}(0) = \tilde{f}_{A}.
\end{gather}
In each of the four cases the coefficients $\sigma_{\nu}$ can be
uniquely determined, correspondingly, by
\begin{gather}
\label{eq40}
\sigma_{1}=\left\{ \frac{r}{\sqrt{s^2+|\omega|^2}+\theta
s}\right\}\tilde{g},\
\sigma_{2,3}=\left\{ \frac{r}{\theta \sqrt{s^2+|\omega|^2}+s}\right\}
[\tilde{g}_{2,3}-\frac{\theta\imath\omega_{2,3}}{\sqrt{s^2+|\omega|^2}+\theta s}\tilde{g}];\\
\label{eq41}
\sigma_{1}=\left\{ -\frac{r}{\sqrt{s^2+|\omega|^2}}\right\}\tilde{h},\quad
\sigma_{2,3}=\left\{ \frac{r}{s}\right\}
\tilde{h}_{2,3};\\
\label{eq42}
\sigma_{1}=\left\{ \frac{r}{s}\right\}\tilde{h},\quad
\sigma_{2,3}=\left\{-\frac{r}{\sqrt{s^2+|\omega|^2}}\right\}\tilde{h}_{2,3};\\
\label{eq43}
\sigma_{1}=\left\{ \frac{r}{\sqrt{s^2+|\omega|^2}(\sqrt{s^2+|\omega|^2}+s)}\right\}\imath\omega^{A}\tilde{f}_{A},\quad
\sigma_{2,3}=\left\{\frac{r}{\sqrt{s^2+|\omega|^2}+s}\right\}\tilde{f}_{2,3}.
\end{gather}
Once the coefficients $\sigma_{\nu}$ are found, the solution to
(\ref{eq30}) is obtained by substituting (\ref{eq38}) into
(\ref{eq35}).

The necessary condition for stability is that (\ref{eq30}) must
admit only the trivial solution satisfying the homogeneous boundary
conditions \cite{GKO95,KL89,KW06}. Otherwise, with each solution
(\ref{eq35}), the function
\begin{equation*}
\left(\begin{array}{c}
\pi_{i} \\
\varphi_{ji}
\end{array}
\right) =
\left(\begin{array}{c}
\tilde{\pi}_{i}(a x_{1}) \\
\tilde{\varphi}_{ji}(a x_{1})
\end{array}
\right)\exp(ast+\imath a\omega_{A}x^{A})
\end{equation*}
is also a solution, for any $a>0$. Since we can choose the parameter
$a$ to be arbitrarily large, the solution of (\ref{eq30}) can not be
bounded by any exponential function \cite{KL89}. The corresponding
initial-boundary value problem is therefore strongly
unstable.

It can be seen from (\ref{eq40})--(\ref{eq43}) that if
$\tilde{g}_{A}=\tilde{h}_{A}=\tilde{f}_{A}=0$,
$\tilde{g}=\tilde{h}=0$, then the coefficients $\sigma_{\nu}=0$.
Thus $(\tilde{\pi}_{i}(x_{1}),\tilde{\varphi}_{xi}(x_1))\equiv 0$,
and the necessary condition for stability is satisfied for each of
the four sets of boundary conditions.

In the case of vanishing initial data, well-posedness of the
initial-boundary value problem for system (\ref{eq30}) follows if
(\ref{eq38}) is uniformly bounded by the boundary
data. This is also known as the Kreiss condition
\cite{GKO95,KL89,KW06}. Specifically, if a constant $K>0$ independent of
$s$ and $\omega_{A}$ can be found such that (here $\| \cdot \|$ is the pointwise vector norm)
\begin{equation}
\label{eq44} \| (\tilde{\pi}_{i}(0), \tilde{\varphi}_{xi}(0)) \|
\le K  \| \tilde{g}, \tilde{g}_{A} \| \quad (\le K \| \tilde{h}, \tilde{h}_{A} \|, \quad \le K \|
\tilde{f}_{A} \|),
\end{equation}
then the unique solution to (\ref{eq30}) can be constructed by the
inverse Laplace-Fourier transform, and can also be estimated by the
boundary data \cite{GKO95,KL89}.

It can be noticed that Kreiss condition is not satisfied by
(\ref{eq32}) and (\ref{eq33}). Indeed, by fixing $\omega_{A}$ and
letting $s\to 0$, one can use (\ref{eq41}), (\ref{eq42}) to verify
that the coefficients $\sigma_{\nu}$ can not be bounded uniformly
in $s$. Also, the coefficients can not be uniformly bounded
in $\omega_{A}$ either.

To discuss (\ref{eq31}) and (\ref{eq34}) we will need the
following lemma from \cite{RS05}
\begin{lemma}
\label{lem5}
Let $P>0$, $A,B\in \R$, $(A,B)\neq (0,0)$, and consider the
function
\begin{equation*}
\psi:\{ \Re(\zeta)>0\} \to \C, \quad
\psi(\zeta)=A\zeta-B\sqrt{\zeta^2+P^2},
\end{equation*}
where we choose the branch of the square root such that $\Re( \sqrt{\zeta^2+P^{2}})>0$ for
$\Re(\zeta)>0$.

Then, $\psi$ has zeros if and only if $A>B>0$ or $A<B<0$.
Furthermore, $|\psi|$ is uniformly bounded away from zero if and only if
$AB<0$.
\end{lemma}

From Lemma~{\ref{lem5}} it is apparent that as long as
$|\alpha|<1$, both $\sqrt{s^2+|\omega|^2}+\theta s$ and
$\theta\sqrt{s^2+|\omega|^2}+s$ are uniformly bounded away from
zero. This implies that coefficients $\sigma_{\nu}$  in
(\ref{eq40}) remain bounded if $s\to 0$, $\Re(s)>0$. At the same
time, when $|s|\to \infty$,
\begin{equation*}
\frac{r}{|\sqrt{s^2+|\omega|^2}+\theta s|}=
\frac{\sqrt{|s^2+|\omega|^2|+|s|^2}}{|\sqrt{s^2+|\omega|^2}+\theta s|}=
\frac{\sqrt{|1+|\omega|^2/s^2|+1}}{|\sqrt{1+|\omega|^2/s^2}+\theta
|}\quad  \longrightarrow_{\hspace*{-6mm}\rule{0mm}{5mm}|s|\to \infty} \quad
\frac{2}{1+\theta}.
\end{equation*}
Similarly, when $|\omega|^{2}\to \infty$, the above expression
converges to $1$. Therefore, we conclude that $\sigma_{1}$ is
uniformly bounded by $\tilde{g}$. By a similar argument,
$\sigma_{2,3}$ are also uniformly bounded by $\tilde{g}_{2,3}$ and
$\tilde{g}$ since the coefficient in braces can be treated similar
to that of $\sigma_{1}$, and also, $|\omega|
(\sqrt{s^2+|\omega|^2}+\theta s)^{-1}\to 1$ as $|\omega|\to
\infty$. Thus the Kreiss condition is satisfied for (\ref{eq31}).

The condition (\ref{eq34}) does not meet the Kreiss condition
since $\sigma_{1}$ in (\ref{eq43}) depends on
$1/\sqrt{s^2+|\omega|^2}$, which can be made large by choosing
$s=|\varepsilon|+i|\omega|$, $\varepsilon\to 0$. However, the
coefficients remain bounded in the high frequency regime, when
$|\omega|\to \infty$.

\begin{remark}
The necessary condition for stability is equivalent to the
non-vanishing determinant of the corresponding matrix for the
coefficients $\sigma_{\nu}$ (cf., \cite{GKO95,KL89,KW06,R06}). At
the same time, for the Kreiss condition the explicit formulae for
the coefficients $\sigma_{\nu}$ are more informative. In
particular, in (\ref{eq40}), the matrix appears to
be lower-triangular and thus its determinant does not include
off-diagonal terms. As the result, terms
$(\theta\imath\omega_{2,3})(\sqrt{s^2+|\omega|^2}+\theta s)^{-1}$
in the expressions for $\sigma_{2,3}$ are overlooked in the
determinant condition. However, they clearly depend on $\omega_{A}$.
\end{remark}

\section{Weak instabilities}

A framework to quantify instabilities that grow polynomially in
time was formulated in \cite{R06}. Such instabilities may be
triggered by small perturbations of the initial data, and, for certain
types of boundary conditions, can grow faster in time than a
polynomial of any degree. We complement the analysis of Section~7
with this technique to see if non-vanishing initial data can
produce a milder instability, similar to the one described in
\cite{RS05}.

Following \cite{R06}, we perform the Fourier transform of the
solution in tangential variables $x_{A}$ and look for a solution in the form
\begin{equation}
\label{eq45}
\left(\begin{array}{c}
\pi_{i} \\
\varphi_{ji}
\end{array}
\right) =
\exp(\imath\omega\omega_{A}x^{A})
\sum_{\nu=0}^{p}
\left(\begin{array}{c}
\tilde{\pi}^{(\nu)}_{i}(\omega x_{1}) \\
\tilde{\varphi}^{(\nu)}_{ji}(\omega x_{1})
\end{array}
\right)
\frac{(\omega t)^{\nu}}{\nu !},
\end{equation}
where $\omega\in \R$, $\omega>0$, $\omega_{A}$ is a real
two-vector, $(\tilde{\pi}^{(p)}_{i},
\tilde{\varphi}^{(p)}_{ji})\neq 0$, and $p\ge 1$.

Substituting (\ref{eq45}) into (\ref{eq30}), and grouping terms by powers of $t$, we have\\
$t^{p}$:
\begin{align}
\label{eq45a}
0&=\partial_{x}\tilde{\varphi}^{(p)}_{xi}(\omega x_{1})+\imath \omega^{A}\tilde{\varphi}^{(p)}_{Ai} (\omega x_{1}),\\
0&=\partial_{x}\tilde{\pi}^{(p)}_{i}(\omega x_{1}),\nonumber \\
\label{eq46}
0 & = \imath \omega_{A} \tilde{\pi}^{(p)}_{i}(\omega x_{1});
\end{align}
$t^{\nu}$, $\nu=0,\ldots,p-1$:
\begin{align}
\label{eq47}
\tilde{\pi}^{(\nu+1)}_{i}(\omega x_{1}) & =\partial_{x}\tilde{\varphi}^{(\nu)}_{xi}(\omega x_{1})+\imath
\omega^{A}\tilde{\varphi}^{(\nu)}_{Ai} (\omega x_{1}), \\
\label{eq48}
\tilde{\varphi}^{(\nu+1)}_{xi}(\omega x_{1})&=\partial_{x}\tilde{\pi}^{(\nu)}_{i}(\omega x_{1}),\\
\label{eq49}
\tilde{\varphi}^{(\nu+1)}_{Ai}(\omega x_{1}) &=\imath \omega_{A} \tilde{\pi}^{(\nu)}_{i}(\omega x_{1}).
\end{align}

Obviously, $\tilde{\pi}^{(p)}_{i}\equiv 0$ as long as
$\omega_{A}\ne 0$. We let $\nu=p-1$, differentiate (\ref{eq48}) in
$x_{1}$, and substitute the result in (\ref{eq45a}). Similarly, we
substitute (\ref{eq49}) in (\ref{eq45a}) to obtain
\begin{equation*}
0=\partial^2_{x} \tilde{\pi}^{(p-1)}_{i}(\omega x_{1}) -|\omega_{A}|^2\tilde{\pi}_{i}^{(p-1)}(\omega x_{1}).
\end{equation*}
The most general solution that is bounded at infinity is
\begin{equation*}
 \tilde{\pi}^{(p-1)}_{i}(\omega x_{1}) =
 \sum_{\alpha=1}^{3}\sigma^{(p-1)}_{\alpha}v_{\alpha}\exp(-|\omega_{A}|x_{1}),
\end{equation*}
where $v_{1}=(1,0,0)$, $v_{2}=(0,1,0)$, and $v_{3}=(0,0,1)$.

The coefficients $\sigma^{(p-1)}_{\alpha}$ are to be determined
from the boundary conditions in view of the following relations
that are consequences of (\ref{eq48}) and (\ref{eq49})
\begin{equation}
\label{eq50}
\tilde{\varphi}^{(p)}_{xi}(\omega x_{1})= -|\omega_{A}| \tilde{\pi}^{(p-1)}_{i}(\omega
x_{1}),\quad
\tilde{\varphi}^{(p)}_{Ai}(\omega x_{1})= \imath\omega_{A} \tilde{\pi}^{(p-1)}_{i}(\omega
x_{1}).
\end{equation}
One can check that $\tilde{\varphi}^{(p)}_{xi}(\omega x_{1})$ and
$\tilde{\varphi}^{(p)}_{Ai}(\omega x_{1})$ obtained from
(\ref{eq50}) satisfy (\ref{eq45a}) automatically.

Similarly, for $\nu=2,\ldots,p$, one finds
\begin{equation*}
\tilde{\pi}^{(\nu)}_{i}(\omega x_{1})=\partial^2_{x} \tilde{\pi}^{(\nu-2)}_{i}(\omega x_{1}) -|\omega_{A}|^2\tilde{\pi}_{i}^{(\nu-2)}(\omega x_{1}),
\end{equation*}
which can be recursively solved. As in the previous case, the general
formulas for $\tilde{\varphi}^{(\nu)}_{xi}$ and $\tilde{\varphi}^{(\nu)}_{Ai}$ can
be obtained from (\ref{eq48}) and (\ref{eq49}), and the coefficients
are determined from the boundary conditions.

Consider the case of boundary conditions (\ref{eq31}). By taking
the Fourier transform in $x_{A}$ of $g_{A}$ and $g$, and  decomposing
the result in the Taylor series, we have
\begin{equation}
\label{eq51}
\theta \tilde{\pi}^{(\nu)}_{x}(0) - \tilde{\varphi}^{(\nu)}_{xx}(0)=\tilde{g}^{(\nu)},
\quad
\tilde{\pi}^{(\nu)}_{A}(0) - \theta(\tilde{\varphi}^{(\nu)}_{xA}(0)- \tilde{\varphi}^{(\nu)}_{Ax}(0))=
\tilde{g}^{(\nu)}_{A},
\end{equation}
where $\nu=0,1,\ldots,p$, and $\theta=\frac{1+\alpha}{1-\alpha}$,
$|\alpha|<1$.

Let functions $g_{A}$ and $g$ be such that
$\tilde{g}^{(p)}_{A}=0$, $\tilde{g}^{(p)}=0$. If a nontrivial
solution (\ref{eq45}) (with some $\tilde{\varphi}^{(p)}_{ki}\neq
0$) can be found satisfying boundary conditions (\ref{eq51}), then
the initial-boundary value problem is ill-posed. Indeed, in this
case, the solution at times $t>0$ is $O(\omega^{p})$ whereas the
initial data is $O(\omega^{0})$ and the boundary data is
$O(\omega^{p-1})$. Therefore, the solution can not be estimated in
terms of the initial data as $\omega\to \infty$, and the problem
is ill-posed \cite{R06}.

Using $\tilde{\pi}^{(p)}_{i}\equiv 0$ and
(\ref{eq50}) we have
\begin{equation}
\sigma^{(p-1)}_{1} = \frac{1}{|\omega_{A}|}\tilde{g}^{(p)},\quad
\sigma^{(p-1)}_{2,3}=\frac{1}{\theta |\omega_{A}|}
\tilde{g}^{(p)}_{2,3}-\frac{\imath \omega_{2,3}}{|\omega_{A}|^2}
\tilde{g}^{(p)}.
\end{equation}
This guarantees that $\tilde{\pi}^{(p-1)}_{i}(\omega x_{1})$, and
thus $\tilde{\varphi}^{(p)}_{xi}(\omega x_{1})$ and
$\tilde{\varphi}^{(p)}_{Ai}(\omega x_{1})$ are zero as long as
$\tilde{g}^{(p)}_{A}=0$ and $\tilde{g}^{(p)}=0$. Therefore, (\ref{eq31})
is polynomially stable.

At the same time boundary conditions (\ref{eq32}) yield,
\begin{equation*}
\tilde{\varphi}^{(\nu)}_{xx}(0)=\tilde{h}^{(\nu)},\quad
\tilde{\pi}^{(\nu)}_{A}(0)=\tilde{h}^{(\nu)}_{A},\qquad
\nu=0,1,\ldots,p.
\end{equation*}
It can be seen that if $\tilde{h}^{(p)}_{A}=0$ and
$\tilde{h}^{(p)}=0$, but $\tilde{h}^{(p-1)}_{A}\neq 0$, the second
condition implies that $\tilde{\pi}^{(p-1)}_{A}(\omega x_{1}) =
\tilde{h}^{(p-1)}_{A} \exp(-|\omega_{A}|x_{1}) \neq 0$, and, in
view of (\ref{eq50}), $\tilde{\varphi}^{(p)}_{xA}(\omega x_{1})=
-|\omega_{A}|\tilde{h}^{(p-1)}_{A} \exp(-|\omega_{A}|x_{1}) \neq
0$. Thus condition (\ref{eq32}) admits ill-posed polynomially
growing modes (unless $h_{A}=0$, $h=0$).

Similarly, (\ref{eq33}) is ill-posed, since it implies that if
$\tilde{h}^{(p-1)}\neq 0$, then
$\tilde{\varphi}^{(p)}_{xx}(\omega x_{1})= -|\omega_{A}|
\tilde{h}^{(p-1)} \exp(-|\omega_{A}|x_{1}) \neq 0$.

Finally, consider boundary conditions (\ref{eq34}). Using the Fourier
transformation in variables $x_{A}$ and substituting decomposition
(\ref{eq45}), the boundary condition yields
\begin{gather}
\label{eq52}
\tilde{\pi}^{(\nu)}_{A}(0) - \tilde{\varphi}^{(\nu)}_{xA}(0) = \tilde{f}^{(\nu)}_{A},\quad
\nu=0,1,\ldots,p,\\
\label{eq53}
\omega^{A}(\tilde{f}^{(p)}_{A}+\tilde{\varphi}^{(p)}_{Ax}(0))=0,\\
\label{eq54}
\tilde{\pi}^{(\nu)}_{x}(0)-\tilde{\varphi}^{(\nu)}_{xx}(0)-
\imath\omega^{A}(\tilde{f}^{(\nu-1)}_{A}+\tilde{\varphi}^{(\nu-1)}_{Ax}(0))=0,\quad
\nu=1,2,\ldots,p.
\end{gather}
Since $\tilde{\pi}^{(p)}_{i}\equiv 0$, from (\ref{eq52}) it
follows that $\tilde{\varphi}^{(p)}_{xA}=0$ if
$\tilde{f}^{(p)}_{A}=0$. Therefore, from (\ref{eq50}),
$\tilde{\pi}^{(p-1)}_{A}=0$ and $\tilde{\varphi}^{(p)}_{BA}=0$. It
remains to check that $\tilde{\varphi}^{(p)}_{xx}=0$, and
$\tilde{\varphi}^{(p)}_{Ax}=0$. First of all
$\omega^{A}\tilde{\varphi}^{(p)}_{Ax}=0$ due to (\ref{eq53}).
This, in view of (\ref{eq50}), implies that
$\tilde{\pi}^{(p-1)}_{x}=0$ and $\tilde{\varphi}^{(p)}_{xx}=0$.
By (\ref{eq50}) this yields $\tilde{\varphi}^{(p)}_{Ax}=0$
as well. Thus boundary condition (\ref{eq34}) is stable in the
sense that no polynomially growing modes that can not be
estimated by the data is generated.

\section{Acknowledgements}

The author thanks O.~Rinne for collaboration and fruitful
discussions, and the Caltech Numerical Relativity group for
helpful comments and interest in this work. The author thanks
O.~Sarbach for suggesting improvements to results of Section~5.
The author thanks L.~Lindblom, M.~Scheel, H.~Pfeiffer, D.~Arnold,
and L.~L.~Foster for illuminating conversations, support, and
helpful suggestions that helped this work to appear. The author
thanks the JHDE reviewers for their meticulous work and
consequential comments that resulted in many improvements to this
manuscript.

This work was supported by the Probationary Faculty
Research Support Program at California State University,
Northridge.

\bibliographystyle{plain}
\bibliography{n031407}
\end{document}